\documentclass{emulateapj}
\usepackage{apjfonts} 
\slugcomment{Submitted to ApJL on 25 February 2015; accepted on 13 March 2015}

\begin{document}

\shortauthors{ALMA Partnership et al.}

\shorttitle{ALMA Observations of Juno}

\title{ALMA Observations of Asteroid 3 Juno at 60 Kilometer Resolution}

\author{ALMA Partnership, 
T.~R.~Hunter\altaffilmark{1}, 
R.~Kneissl\altaffilmark{2,3}, 
A. Moullet\altaffilmark{1}, 
C.~L.~Brogan\altaffilmark{1}, 
E. B. Fomalont\altaffilmark{2,1},
C. Vlahakis\altaffilmark{2,3},
Y. Asaki \altaffilmark{4,5},
D. Barkats\altaffilmark{2,3},  
W. R. F. Dent\altaffilmark{2,3},
R. E. Hills\altaffilmark{6},
A. Hirota\altaffilmark{2,4},  
J. A. Hodge\altaffilmark{1},
C. M. V. Impellizzeri\altaffilmark{2,1},
E. Liuzzo\altaffilmark{7},
R. Lucas\altaffilmark{8},
N. Marcelino\altaffilmark{7},
S. Matsushita\altaffilmark{9},
K. Nakanishi\altaffilmark{2,4},  
L. M. P\'erez\altaffilmark{10},
N. Phillips\altaffilmark{2,3}, 
A. M. S. Richards\altaffilmark{11},
I. Toledo\altaffilmark{2},
R. Aladro\altaffilmark{3},
D. Broguiere\altaffilmark{12}, 
J. R. Cortes\altaffilmark{2,1},
P. C. Cortes\altaffilmark{2,1},
D. Espada\altaffilmark{2,4},
F. Galarza\altaffilmark{2},
D. Garcia-Appadoo\altaffilmark{2,3}, 
L. Guzman-Ramirez\altaffilmark{3},  
A. S. Hales\altaffilmark{2,1},
E. M. Humphreys\altaffilmark{13},  
T. Jung\altaffilmark{14},   
S. Kameno\altaffilmark{2,4}, 
R. A. Laing\altaffilmark{13},     
S. Leon\altaffilmark{2,3},
G. Marconi\altaffilmark{2,3}, 
A. Mignano\altaffilmark{7}, 
B. Nikolic\altaffilmark{6},
L. -A. Nyman\altaffilmark{2,3}, 
M. Radiszcz\altaffilmark{2}, 
A. Remijan\altaffilmark{2,1},
J. A. Rod\'on\altaffilmark{3},  
T. Sawada\altaffilmark{2,4},
S. Takahashi\altaffilmark{2,4},
R. P. J. Tilanus\altaffilmark{15},   
B. Vila Vilaro\altaffilmark{2,3}, 
L. C. Watson\altaffilmark{3},
T. Wiklind\altaffilmark{2,3},
I. de Gregorio-Monsalvo\altaffilmark{2,3},
J. Di Francesco\altaffilmark{16},
J. Mangum\altaffilmark{1},
H. Francke\altaffilmark{2},
J. Gallardo\altaffilmark{2},
J. Garcia\altaffilmark{2},
S. Gonzalez\altaffilmark{2}, 
T. Hill\altaffilmark{2,3},
T. Kaminski\altaffilmark{3}, 
Y. Kurono\altaffilmark{2,4},
C. Lopez\altaffilmark{2},
F. Morales\altaffilmark{2},
K. Plarre\altaffilmark{2},
S. Randall\altaffilmark{13},
T. van kempen\altaffilmark{15}, 
L. Videla\altaffilmark{2},
E. Villard\altaffilmark{2,3},
P. Andreani\altaffilmark{13},
J. E. Hibbard\altaffilmark{1},
K. Tatematsu\altaffilmark{4}
}

\email{thunter@nrao.edu}

% NRAO Cville
\altaffiltext{1}
{National Radio Astronomy Observatory, 520 Edgemont Rd, Charlottesville, VA, 22903, USA}

% JAO
\altaffiltext{2}
{Joint ALMA Observatory, Alonso de C\'ordova 3107, Vitacura, Santiago, Chile}

% ESO Chile
\altaffiltext{3}
{European Southern Observatory, Alonso de C\'ordova 3107, Vitacura, Santiago, Chile}

% NAOJ
\altaffiltext{4}
{National Astronomical Observatory of Japan, 2-21-1 Osawa, Mitaka, Tokyo 181-8588, Japan}

% JAXA/ISAS
\altaffiltext{5}
{Institute of Space and Astronautical Science (ISAS), Japan Aerospace Exploration Agency (JAXA), 3-1-1 Yoshinodai, Chuo-ku, Sagamihara, Kanagawa 252-5210 Japan}

% Cambridge
\altaffiltext{6}
{Astrophysics Group, Cavendish Laboratory, JJ Thomson Avenue, Cambridge, CB3 0HE, UK}

% Bologna
\altaffiltext{7}
{INAF, Istituto di Radioastronomia, Via P. Gobetti 101, 40129, Bologna, Italy}

% Robert Lucas
\altaffiltext{8}
{Institut de Plan\'etologie et d'Astrophysique de Grenoble (UMR 5274), BP 53, 38041, Grenoble Cedex 9, France}
	
% ASIAA
\altaffiltext{9}
{Institute of Astronomy and Astrophysics, Academia Sinica, P.O. Box 23-141, Taipei 106, Taiwan}

% NRAO Socorro
\altaffiltext{10}
{National Radio Astronomy Observatory, P.O. Box O, Socorro, NM 87801, USA}

% Manchester
\altaffiltext{11}
{Jodrell Bank Centre for Astrophysics, School of Physics and Astronomy, University of Manchester, Oxford, Road, Manchester M13 9PL, UK}

% IRAM
\altaffiltext{12}
{IRAM, 300 rue de la piscine 38400 St Martin d'H\`eres, France}

% ESO Garching
\altaffiltext{13}
{European Southern Observatory, Karl-Schwarzschild-Str. 2, D-85748 Garching bei M\"unchen, Germany}

% KASI
\altaffiltext{14}{
Korea Astronomy and Space Science Institute, Daedeokdae-ro 776, Yuseong-gu, Daejeon 305-349, Korea}

% Leiden
\altaffiltext{15}
{Leiden Observatory, Leiden University, P.O. Box 9513, 2300 RA Leiden, The Netherlands}

% NRC NAASC
\altaffiltext{16}
{National Research Council Herzberg Astronomy \& Astrophysics, 5071 West Saanich Road, Victoria, BC V9E 2E7, Canada}

\begin{abstract} % currently at the 250 word maximum

We present Atacama Large Millimeter/submillimeter Array (ALMA) 1.3~mm
continuum images of the asteroid 3 Juno obtained with an angular
resolution of $0.042\arcsec\/$ (60~km at 1.97~AU).  The data were
obtained over a single 4.4~hr interval, which covers 60\% of the
7.2~hr rotation period, approximately centered on local transit.  A
sequence of ten consecutive images reveals continuous changes in the
asteroid's profile and apparent shape, in good agreement with the sky
projection of the three-dimensional model of the Database of Asteroid
Models from Inversion Techniques.  We measure a geometric mean
diameter of 259$\pm$4~km, in good agreement with past estimates from a
variety of techniques and wavelengths.
% \citet{Altenhoff94} size estimate inferred from the single-dish 1.2~mm
% flux density combined with unity emissivity.  
Due to the viewing angle and inclination of the rotational pole, the
southern hemisphere dominates all of the images.  The median peak
brightness temperature is 215$\pm$13~K, while the median over the
whole surface is $197\pm15$~K.  With the unprecedented resolution of
ALMA, we find that the brightness temperature varies across the
surface with higher values correlated to the subsolar point and
afternoon areas, and lower values beyond the evening terminator.  The
dominance of the subsolar point is accentuated in the final four
images, suggesting a reduction in the thermal inertia of the regolith
at the corresponding longitudes, which are possibly correlated to the
location of the putative large impact crater.
%The absolute position
%of Juno's centroid differs by $\approx0.06\arcsec\/$ from 
%existing ephemerides, similar to their quoted $3~\sigma$ level of
%accuracy.  
These results demonstrate ALMA's potential to resolve
thermal emission from the surface of main belt asteroids, and to
measure accurately their position, geometric shape, rotational period,
and soil characteristics.

\end{abstract}


\section{Introduction}

Discovered in 1804, Juno was the third main-belt asteroid identified,
following Ceres and Pallas. % \citep{Cunningham04}.
% but is more than an order of magnitude less massive. 
The first reasonably accurate measurement of Juno's diameter was
performed with filar micrometers on the Great Lick Refractor
\citep{Barnard95} and the Yerkes 40-inch Refractor \citep{Barnard00},
yielding a value of $193\pm20$~km \citep[see also][]{Dollfus71}.  A
modern measurement of its physical cross section came from 15-station
stellar occultation data, yielding a mean diameter of 267$\pm$5~km
with a significant ellipticity \citep{Millis81}.  Optical speckle
interferometry soon produced a size measurement consistent with the
occultation result \citep{Baier83}.  Like most asteroids, Juno's light
curve is double-peaked with two maxima and two minima
\citep[e.g.,][]{Birch89}, indicative of a non-spherical shape. Based
on light curve inversion, Juno has a unique rotational pole that is
significantly tilted with respect to the ecliptic
\citep{Magnusson86,Dotto95}, and its period of 7.209531~hr is known to
high accuracy \citep{Kaas02}.  This information, combined with recent,
near-infrared adaptive optics (AO) imaging led to a triaxial ellipsoid
model with axis lengths of 297, 233, and 222~km
\citep{Drummond08}.  A three-dimensional model with 2036 faces and
1020 vertices based on a combination of the historical optical light
curves and two occultations \citep{Durech11,Kaas02} is hosted by the
Database of Asteroid Models from Inversion Techniques
\citep[DAMIT;][]{Durech10}.

Juno is a member of the S-class of asteroids \citep{Chapman75}, which
have a stony composition of iron-bearing silicates and metallic iron
as inferred primarily from their 1~$\mu$m spectral absorption feature
\citep[e.g.,][and references therein]{Gaffey93a}.  
% Feierberg82,
The optical Small Main-Belt Asteroid Spectroscopic Survey (SMASSII)
assigns it subclass Sk as a transition object toward the K-class,
which exhibits a shallower 1~$\mu$m feature \citep{Bus02}.
% , while \citet{Gaffey93b} assigns it subclass S(IV).
{\it Infrared Space Observatory} spectra of Juno show an 8-11.5~$\mu$m
feature that is consistent with the laboratory measurements of the
silicate minerals pyroxine and olivine \citep{Dotto00}.  
%It has been
%posited as a likely parent body of the ordinary chondrite meteorites
%\citep{Gaffey93b}.
%Arecibo radar
% measurements of its reflectivity and circular polarization ratio are
% typical of S-class asteroids \citep{Magri07}.  
Evidence for surface features on Juno have been suggested by the
variation as a function of rotation angle of its optical colors
\citep{Degewij79,Schroll81} and linear polarization
\citep{Shinokawa02,Takahashi09}, and by a sequence of optical AO
images, which suggested a large impact crater \citep{Baliunas03}.
Somewhat surprisingly, there are no published images of Juno from the
{\it Hubble Space Telescope} \citep[][]{Dotto02}, and
% it does not appear in the {\it WISE}/NEOWISE
% survey of 2835 main-belt asteroids \citep{Masiero14}.  
there have been no spacecraft encounters as yet.

As a powerful new tool in the study of Solar System bodies, the
Atacama Large Millimeter/submillimeter Array \citep[ALMA;][]{Hills10} 
will be able to map the shape and surface temperature distributions of
hundreds of main belt asteroids and Jupiter Trojans
\citep{Busch09,Lovell08}.  The reason is twofold.  First, the
absorption length of (sub)millimeter photons \citep{Campbell69} on
asteroid surfaces is comparable to the thermal skin depth of the
diurnal wave \citep[typically a few to 10 wavelengths;][]{Spencer89};
thus, these wavelengths are well-matched to probe the thermal response
of this material and should provide information on the thickness,
structure and nature of the regolith \citep{Chamberlain07,Lagerros96}.
Indeed, the recent flybys of the main belt asteroids 21~Lutetia and
2867~{\v S}teins by the {\it Rosetta} mission have yielded important
measurements of the thermal inertia and emissivity of their surface
material (\S~\ref{discuss}) using its (sub)millimeter radiometers
\citep{Gulkis10,Gulkis12}.  Second, ALMA has exquisite continuum
brightness temperature sensitivity at small angular scales.  For
example, the 50-antenna, full-bandwidth sensitivity in one minute at
300~GHz (1~mm) at the highest angular resolution (13 mas) is 10~K.
Since their physical temperatures are typically 100-200~K, ALMA can
effectively image these bodies down to linear resolutions of 10~km (at
a distance $\Delta=1$~AU) with high signal-to-noise ratio (S/N),
enabling the use of the powerful tool of self-calibration.  As a
significant step toward demonstrating this capability, in this Letter
we present the first millimeter wavelength images to resolve the
surface of Juno, which were obtained during the recent campaign to
commission ALMA's long baseline capabilities \citep{ALMA15a}.  The
availability of the DAMIT model provides an excellent test of ALMA's
imaging performance, while the resulting images provide new details on
the surface conditions of Juno.

% The largest S-class asteroid visited by a spacecraft is Ida (56x24 km).

\section{Observations}

An approximately 53-minute length scheduling block (SB) to observe the
1.3~mm (233~GHz) continuum emission from Juno was executed five consecutive
times on 2014 Oct 19 starting at 09:15 UT (43~min before local
sunrise) and ending at 13:38~UT.  Four spectral windows were used,
each with 2~GHz bandwidth, 128 channels and dual polarization. 
Center frequencies were 224, 226, 240, and 242~GHz.
All necessary calibration observations were
performed in each execution of the SB. An additional focus measurement
and adjustment was performed prior to the fifth execution (two hours
after sunrise)
% when the Sun had reached an elevation of 38\arcdeg\/ 
as per normal operations.  The SB included an external ephemeris with
4~min sampling obtained from Jet Propulsion Lab (JPL) 
Horizons\footnote{\url{http://ssd.jpl.nasa.gov/horizons.cgi}},
%\citep{Giorgini96}, 
which reports a $3~\sigma$ uncertainty of 60~mas in
right ascension and 26~mas in declination.

Calibration and imaging was performed in
CASA\footnote{\url{http://casa.nrao.edu}} version 4.2.2.  The complex
gain calibration cycle time was 68~s, with Juno being observed for
48~s and the gain calibrator (J0757+0956) for 15~s (5.7\arcdeg\/
away).
%This switching cycle is more rapid than is typically used for
%ALMA science observations in smaller configurations. 
Data from 27 to 31~antennas were used, ranging in projected baseline length
%
% Petrov webpage    = 07 57 06.642950  +09 56 34.85226
% Lanyi 2010 Q-band = 07 57 06.642955  +09 56 34.85237
% Lanyi 2010 K-band = 07 57 06.642945  +09 56 34.85244
% ALMA catalog      = 07:57:06.642960  +09.56.34.85256
%
from 0.02 to 10~M$\lambda$ (26~m to 13~km).  The zenith precipitable
water vapor varied from 1.4 to 1.6~mm.  Bandpass and flux
calibration is based on observations of the quasar J0750+1231 in each
SB.  This quasar is an ALMA calibration grid source
\citep{vanKempen14} 
for which a linear interpolation in
frequency from the measurements nearest in time at 3~mm (7 days) and
0.87~mm (18 days) yields an assumed flux density of 0.64$\pm$0.04~Jy
at 233~GHz with a spectral index of $-0.66$. The calibrator flux
density measurements were stable over many weeks and we estimate our
flux scale to be accurate to 6\%.  The mean flux density derived for
the gain calibrator was quite consistent across the five executions,
differing by a maximum of 2.4\% from the weighted mean of all
executions ($0.5916\pm0.0007$~Jy). Thus, there was very little
decorrelation on timescales shorter than the integration time
(1.92~s). As shown in Table~\ref{table}, the span of observations was
approximately centered on the time of transit of Juno.

Following calibration, the uv data from each execution was split into
two halves, with the duration and time on source of each half being
$\sim$18~min and $\sim$10~min, respectively.  The ten resulting
datasets (epochs) were imaged individually using a robust weighting 
parameter of zero.  Phase-only self-calibration was then performed,
initially with a time interval of 300~s, followed by a refinement with
a time interval of 15~s.  The final execution showed phase-cal
solutions of somewhat larger magnitude and higher variability than the
rest, with a loss of some antennas on the outermost pads.  Amplitude
self-calibration on a timescale of 300~s was then performed.  The
final images were constructed using multi-scale clean \citep{Rau11}
with deconvolution scales of 0, 5, and 15 pixels to avoid image
artifacts caused by the clean instability that can occur when modeling
an extended source using only delta functions.  The
% Juno0 7.35/0.12=61   7.33/0.065=113
% Juno7 5.36/0.23=23   9.64/.080=121
% Juno9 3.73/0.26=14  14.1/0.114=124
image dynamic range after self-calibration was 120. This improvement
factor of 2-6 indicates that a significant residual phase error was
present after normal calibration.  Thus, it is important to realize
that any usage of calibrated visibilities for direct modeling
\citep[e.g.,][]{Viikinkoski15} must either apply the imaging
self-calibration solutions or include antenna-based phase solutions as
a model parameter to be solved \citep{Hezaveh13}.  As expected when
the initial self-calibration model has high S/N, the
intensity-weighted centroid of the source before and after
self-calibration is consistent to within a fraction of the (5~mas)
pixel size ($<0.2$ in most images, and $<0.3$ in the final two
images).  Because we used an accurate VLBI position for the gain
calibrator \citep[07:57:06.64296, +09:56:34.8525;][]{Lanyi10}, we
expect our images to follow ALMA's measured astrometric performance
\citep[3~mas;][]{ALMA15a}\footnote{The ALMA control system does not
  account for the finite distance of solar system targets when
  computing the gravitational deflection of light by the Sun to apply
  to the astrometric ephemeris, which causes the phase transfer from
  the quasar to introduce a systematic position error of $\approx
  1.4$~mas for these observations of Juno according to the formulas of
  \citet{Cowling84}.}.  The images vary in the size of the synthesized
beam from $31.8\times23.7$~mas to $41.8 \times 36.1$~mas, with a mean
position angle of $30\arcdeg\pm11$\arcdeg.  These images are publicly
available from the ALMA Science Verification
page\footnote{\url{http://almascience.org/alma-data/science-verification.
    Additional details including the calibration and imaging scripts
    are available from
    \url{http://casaguides.nrao.edu/index.php?title=ALMA2014\_LBC\_SVDATA.}}}.
To present matched resolution images, we smoothed the images with a
two dimensional elliptical Gaussian to obtain a circular beam of
42.0~mas, which at the Earth-Juno distance of $\Delta=1.97$~AU
corresponds to 60~km.  No near-field correction was applied to the uv
data, but Juno was beyond the Fraunhofer distance for the longest
baseline, and the self-calibration process may mitigate the residual
effects.

%The expected mean angular diameter was 175~mas, which is about
%half of the maximum size it can reach at a favorable opposition.

% (16.4 light minutes)

\section{RESULTS}

\subsection{Viewing geometry}

\begin{figure*}
\plotone{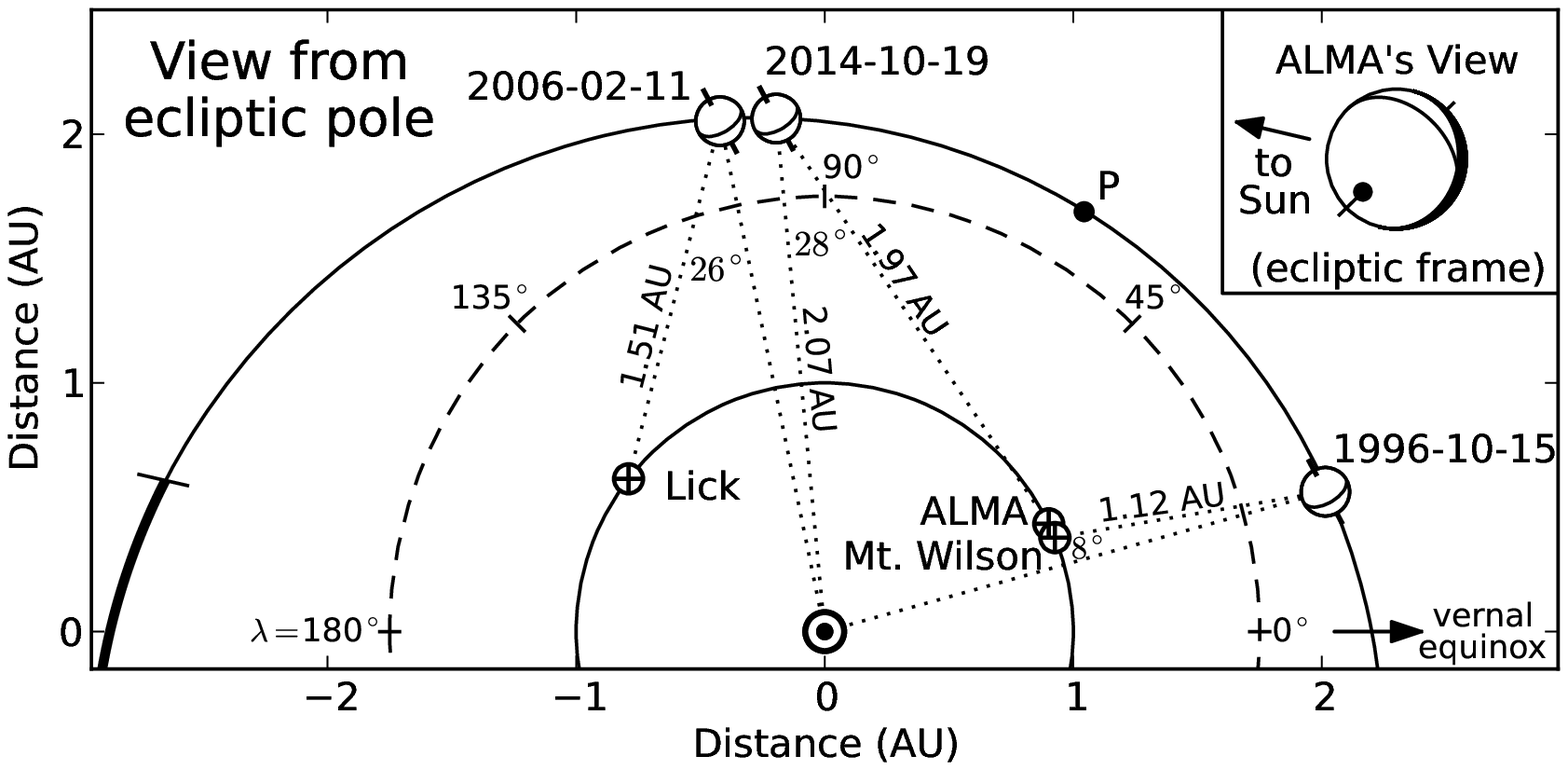}  % diagram.eps
\caption[]{As viewed from the ecliptic pole, this diagram shows the
  alignment of Earth and Juno in their respective orbits on the three
  dates of observational data discussed: ALMA (this paper), Mt. Wilson
  Observatory \citep{Baliunas03}, and Lick Observatory
  \citep{Drummond08}. The scale of heliocentric ecliptic coordinates
  ($\lambda$, $\beta$) is indicated by the dashed line circle.  The
  three angles inside the dotted lines correspond to the
  sun-target-earth angle which describes the solar illumination phase.
  The nominal rotational pole \citep[toward $\lambda=118\arcdeg,
  \beta=+30\arcdeg$;][]{Drummond08} and its corresponding equator are
  drawn onto a spherical representation of Juno for reference.  Juno's
  perihelion at $\lambda=58.3\arcdeg$ is marked by the dot labeled
  ``P''.  The portion of the orbit above the ecliptic is shown by the
  thick line, which begins at the ascending node
  ($\lambda=+169.9\arcdeg$).  {\it Inset:} The inset shows the point
  of view of the ALMA observations (in the ecliptic coordinate frame),
  which is dominated by the southern hemisphere. Juno's south pole and
  equator are marked, as is the evening terminator and unlit side
  which reflects Juno's solar phase angle and heliocentric latitude
  ($\beta = -13.2\arcdeg\/$ viewed from Earth) at the time of
  observation. }
\label{orbitdiagram}
\end{figure*}

Juno's orbit has a mean radius of 2.67~AU with significant
eccentricity (0.25) and inclination (13\arcdeg).  As shown in
Figure~\ref{orbitdiagram}, the true anomaly was $37.5\arcdeg$ during
our observations, with a solar phase angle of -28\arcdeg\/ yielding an
illumination percentage of 94\%.  We have plotted the orientation of
the rotational pole inferred from parametric blind deconvolution of
near-infrared AO images (ecliptic longitude $\lambda = 118\arcdeg\/$
and latitude $\beta=+30\arcdeg\/$) which has an uncertainty of
13\arcdeg\/ \citep{Drummond08}.  The direction of Juno was $\lambda =
124.7\arcdeg, \beta=-13.2\arcdeg$, thus our viewing angle was
48\arcdeg\/ from the polar axis but nearly coplanar, with the southern
hemisphere dominating the view, as shown in the inset of
Figure~\ref{orbitdiagram}.  The angular smearing due to the
15\arcdeg\/ of axial rotation during each 18~min image leads to 21~mas
of linear smearing at the mean radius, which is half the beam size,
leading to $<12$\% loss in resolution.  The mean epochs of the images
are listed in Table~\ref{table}, along with the corresponding
rotational phases computed with respect to the zero time point of the
\citet{Drummond08} triaxial model, and with respect to the optical AO
observations of \citet{Baliunas03}.

\begin{deluxetable}{cccccc}[t] 
\tabletypesize{\footnotesize} \tablewidth{0pc}
\tablecaption{Parameters of of the 1.3~mm ALMA images of Juno}
\tablecolumns{6} \tablehead{ \colhead{Image} & \colhead{Epoch} &
  \colhead{Elapsed time} & \colhead{Elevation} &
  \multicolumn{2}{c}{Rotation phase} \\ 
\colhead{\#} & \colhead{(MJD)} &
  \colhead{(minutes)} & \colhead{(\arcdeg)} &
  \colhead{$\phi_{\rm 1}$\tablenotemark{a}} & 
  \colhead{$\phi_{\rm 2}$\tablenotemark{b}}}
\startdata 
0 & 56949.39167 &   0 & 53.2 & 0.33 & 0.18 \\ % .39142 (exact mean of Juno start/stop) 
1 & 56949.40417 &  18 & 55.7 & 0.37 & 0.22 \\ % .40422 
2 & 56949.42917 &  36 & 59.5 & 0.45 & 0.30 \\ % .42936 
3 & 56949.44167 &  73 & 60.5 & 0.49 & 0.35 \\ % .44219 
4 & 56949.46667 & 110 & 60.5 & 0.58 & 0.43 \\ % .46753 
5 & 56949.48056 & 128 & 59.4 & 0.62 & 0.47 \\ % .48041 
6 & 56949.51111 & 173 & 54.7 & 0.72 & 0.58 \\ % .51175 
7 & 56949.52500 & 192 & 51.9 & 0.77 & 0.62 \\ % .52456 
8 & 56949.54861 & 226 & 46.1 & 0.85 & 0.70 \\ % .54834 
9 & 56949.56111 & 244 & 42.6 & 0.89 & 0.74    % .56119
% mylist = [56949.39167,56949.40417,56949.42917,56949.44167, 56949.46667,
%           56949.48056,56949.51111,56949.52500,56949.54861,56949.56111]
% tt.rotationPhase(53777.100127, .300397125, mylist)
% tt.rotationPhase(50371.24167, .300397125, mylist)
\enddata
\tablenotetext{a}{Phase with respect to $\psi_{\rm
    0}$ of the \citet{Drummond08} triaxial model using the rotation
  period of 0.300397125 days \citep{Kaas02}. In the
  8.5-year interval, the phase accumulated uncertainty is $\pm0.01$.}
\tablenotetext{b}{Phase with respect to AO
  observations of \citet{Baliunas03} (zero point taken to 
  be 50371.24167) using the same rotation  period.  In the 
  18-year interval, the accumulated phase uncertainty is $\pm0.02$.}
%
% 1.97 AU *8.3167464 light minutes/AU = 16.383990407999999 minutes
% 16.4/1440. = 0.011378 days
% (16.4/1440.)/0.3003969 = 0.037912804322843836 rotations

% mylist = [56949.39142,56949.40422,56949.42936,56949.44219,56949.46753,
%           56949.48041,56949.51175,56949.52456,56949.54834,56949.56119]
% tt.rotationPhase(53777.100127, .3003969, mylist)
%
% Drummond08
% CASA <24>: au.dateStringToMJD('2006-02-11 02:24:11')
%MJD= 53777.10013, MJDseconds= 4646341451.0, JD= 2453777.60013

% Baliunas03
%CASA <7>: au.dateStringToMJD('1996-10-15 05:48')
%MJD= 50371.24167, MJDseconds= 4352075280.0, JD= 2450371.74167
%       phase(deg)   minutes   MJD          JD
% image 0: 3          3.6     50371.24417   2450371.74417
% image 1: 21         25.2    50371.25625   2450371.75625
% image 2: 42.5       51.1    50371.27716   2450371.77716
% image 3: 60         72.1    50371.29174   2450371.79174
% image 4: 103       123.8    50371.32764   2450371.82764
% image 5: 120       144.2    50371.34181   2450371.84181
% image 6: 137.5     165.2    50371.35639   2450371.85639
% image 7: 159       191.1    50371.37438   2450371.87438
%
% RA, Dec rates:  [ 0.01452181, -0.00554769] arcsec per sec

\label{table}
\end{deluxetable}

\begin{deluxetable*}{cccccccccc}[t]  
\setlength\tabcolsep{1pt}
\tabletypesize{\footnotesize} 
\tablewidth{6.5in} 
\tablecaption{Derived properties from the 1.3~mm ALMA observations} 
\tablecolumns{10}  
\tablehead{ 
  \colhead{Image} 
  & \multicolumn{2}{c}{Centroid of Juno\tablenotemark{a}}  
  & \colhead{Offset\tablenotemark{b}} 
  & \colhead{Integrated} 
  & \colhead{Peak} 
  & \multicolumn{2}{c}{$T_{\rm B}$} 
  & \multicolumn{1}{c}{$T_{\rm B}$ Peak} 
  & \multicolumn{1}{c}{Elliptical disk model} \\ 
  \colhead{\#} 
  & \multicolumn{2}{c}{Absolute position}   
  & \multicolumn{1}{c}{$\Delta\alpha$, $\Delta\delta$ } 
  & \colhead{Flux density\tablenotemark{c,d}} 
  & \colhead{Intensity\tablenotemark{d,e}}  
  & \colhead{Peak\tablenotemark{d}} & \colhead{Median\tablenotemark{d,f}} & \colhead{Position} 
  & \colhead{Parameters\tablenotemark{g}} \\ % & \colhead{Rotation\tablenotemark{h}} \\
  \colhead{} & \colhead{$\alpha$ (J2000)} & \colhead{$\delta$ (J2000)} 
  & \colhead{mas, mas}  
  & \colhead{mJy} & \colhead{mJy beam$^{-1}$} & \colhead{K} & \colhead{K}  
  & \colhead{mas, mas} & \colhead{mas$\times$mas (\arcdeg)}  %  & \colhead{\arcdeg} 
}
\startdata 
% 0 & au.radecOffsetToRadec('08 14 46.8434  +06 19 08.5590', +59 ,  -5, mas=True)  % 09:24 ephemeris position
% 1 & au.radecOffsetToRadec('08 14 47.89425 +06 19 02.5825', +55 ,  +4, mas=True)  % 09:42
% 2 & au.radecOffsetToRadec('08 14 49.99400 +06 18 50.6180', +52 , +14, mas=True)  % 10:18
% 3 & au.radecOffsetToRadec('08 14 51.0432  +06 18 44.632',  +62 , +13, mas=True)  % 10:36
% 4 & au.radecOffsetToRadec('08 14 53.1407  +06 18 32.651',  +63 , +23, mas=True)  % 11:12
% 5 & au.radecOffsetToRadec('08 14 54.3058  +06 18 25.991',  +63 , +26, mas=True)  % 11:32
% 6 & au.radecOffsetToRadec('08 14 56.8698  +06 18 11.326',  +62 , +31, mas=True)  % 12:16
% 7 & au.radecOffsetToRadec('08 14 58.0359  +06 18 04.655',  +55 , +30, mas=True)  % 12:36
% 8 & au.radecOffsetToRadec('08 15 00.02005 +06 07 53.3075', +62 , +47, mas=True)  % 13:10
% 9 & au.radecOffsetToRadec('08 15 01.0716  +06 17 47.296',  +64 , +43, mas=True)  % 13:28
% Planck brightness temps:
0 & 08 14 46.8474 & +06 19 08.554 & +59, -5  & 199.7$\pm$1.5 & 17.0$\pm$0.09 & 222$\pm$1 & 213 & -9, -5   & 200$\times$157 (+47$\pm$3) \\% & +47 \\ % 09:24
1 & 08 14 47.8979 & +06 19 02.586 & +55, +4  & 198.1$\pm$1.5 & 16.8$\pm$0.09 & 220$\pm$1 & 208 & -20, +21 & 199$\times$159 (+37$\pm$3) \\% & +32 \\ % 09:42
2 & 08 14 49.9975 & +06 18 50.632 & +52, +14 & 199.2$\pm$1.5 & 16.5$\pm$0.09 & 216$\pm$1 & 202 & -22, +36 & 196$\times$165 (+16$\pm$4) \\% &  +2 \\ % 10:18
3 & 08 14 51.0474 & +06 18 44.645 & +62, +13 & 200.6$\pm$1.5 & 16.6$\pm$0.09 & 218$\pm$1 & 200 & +8, -3   & 192$\times$169 (+0$\pm$6)  \\% & -13 \\ % 10:36
4 & 08 14 53.1449 & +06 18 32.674 & +63, +23 & 202.9$\pm$1.5 & 16.3$\pm$0.09 & 214$\pm$1 & 197 & +7, +7   & 191$\times$175 (-47$\pm$8) \\% & -43 \\ % 11:12
5 & 08 14 54.3100 & +06 18 26.017 & +63, +26 & 201.8$\pm$1.4 & 16.2$\pm$0.09 & 213$\pm$1 & 196 & +27, +34 & 194$\times$173 (-67$\pm$6) \\% & -60 \\ % 11:32
6 & 08 14 56.8740 & +06 18 11.357 & +62, +31 & 195.8$\pm$1.5 & 15.7$\pm$0.09 & 207$\pm$1 & 186 & +38, +9  & 200$\times$170 (-99$\pm$4)  \\% & -96 \\ % 12:16
7 & 08 14 58.0396 & +06 18 04.685 & +55, +30 & 196.1$\pm$1.5 & 15.8$\pm$0.09 & 208$\pm$1 & 188 & +40, +10 & 200$\times$169 (-108$\pm$4) \\% & -113 \\ % 12:36
8 & 08 15 00.0242 & +06 17 53.355 & +62, +47 & 188.4$\pm$1.8 & 16.3$\pm$0.11 & 215$\pm$2 & 187 & +33, +13 & 192$\times$167 (-126$\pm$5) \\% & -141 \\ % 13:10
9 & 08 15 01.0759 & +06 17 47.339 & +64, +43 & 181.6$\pm$2.0 & 16.4$\pm$0.12 & 215$\pm$2 & 179 & +31, +2  & 187$\times$168 (-137$\pm$6) \\ % & -156 % 13:28 
Median\tablenotemark{h} & .. & .. & +60.5, +24.5 & 198.7$\pm$2.7 & 16.4$\pm$0.2 & 215$\pm$3 & 197$\pm$9 & .. & 194 $\times$ 169  
% median is +60.5, +24.5
\enddata 

\tablenotetext{a}{See section~\S~\ref{images} for details of the centroid calculation.
The systematic uncertainty is estimated to be 3~mas. All coordinates are in the 
ICRF of the ICRS (Epoch J2000.0).}
\tablenotetext{b}{Offset of centroid with respect to the phase center of
the image, which follows the JPL Horizons ephemeris.}
\tablenotetext{c}{Flux
  density integrated over pixels in the region with intensity $>3~\sigma$, where
  $\sigma$ (the image rms) is calculated in an annulus of radii $0.325\arcsec$ to
  $0.575\arcsec$. The quoted uncertainty is: 
  (Number of independent beams in the region)$^{0.5} \times 3~\sigma$, where $\sigma$ is
  the image rms.}
\tablenotetext{d}{The systematic flux calibration uncertainty of 6\% is not included.}
\tablenotetext{e}{The quoted uncertainty is the image rms.}
\tablenotetext{f}{The median brightness temperature ($T_{\rm B}$) is computed using the
Planck equation with the median pixel intensity in the clean model (within the 
central 0.15\arcsec) and the solid angle per pixel.}
\tablenotetext{g}{Major axis, minor axes and position angle east of north. The fit uncertainty on the major and minor axes is 3~mas.}
\tablenotetext{h}{Where listed, the uncertainty is the median absolute deviation from the median.}
%value in this column is set to the fitted value for image 0
%in the first line and increases in subsequent lines at the rotation rate of Juno.
%This value is plotted as the dotted line in Figure~\ref{juno_shifted}.}
\label{tabletwo}
\end{deluxetable*}

\subsection{Images}
\label{images}

\begin{figure*}
\plotone{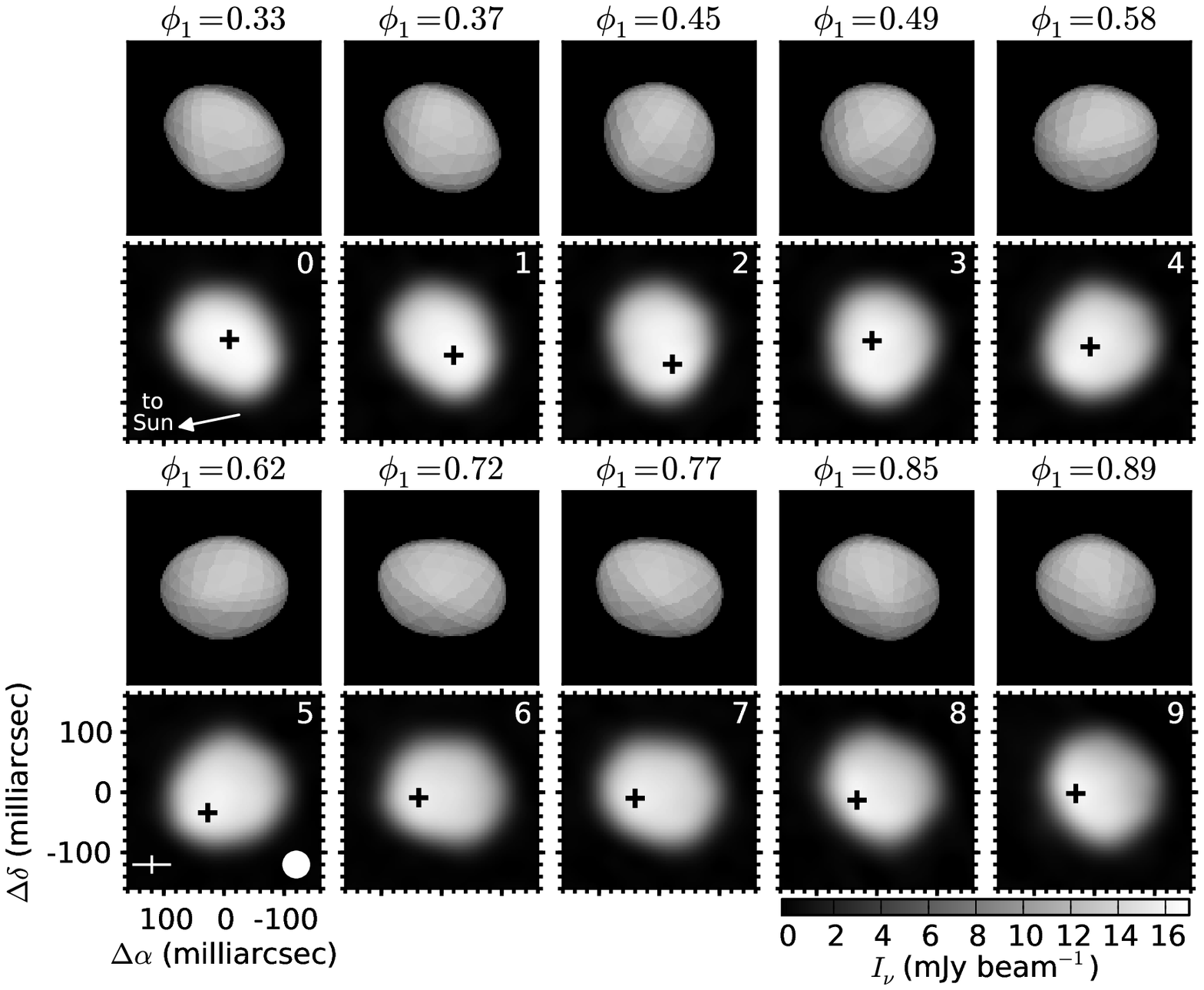} % twentypanel_individual_shifts.eps
\caption[]{In the two pairs of rows, the lower set of panels 
   (with tick marks) are ALMA 1.3~mm continuum images of Juno
  in equatorial offset coordinates referenced to the phase tracking
  center corresponding to the JPL ephemeris, but with the centroid
  shifted by the values in columns 4 and 5 in Table~\ref{tabletwo}. In
  each frame, the cross marks the position of the peak intensity.  The
  position angle of the Sun was +101.5\arcdeg\/ east of north in
  equatorial coordinates, as indicated by the arrow.  
%The dotted line is drawn through the
%  centroid at the position angle of the elliptical model fit to the
%  data only in the first frame -- the one with least uncertainty
%  (Table~\ref{tabletwo}) -- and incremented clockwise at the rotation
%  rate of Juno.  
The beam size is indicated by the circle in the lower
  right-hand corner.  The $3~\sigma$ position uncertainty of the JPL
  ephemeris in each axis is indicated by the cross in the lower
  left-hand corner.  The rotational phase values ($\phi_{\rm 1}$) are
  from Table~\ref{table}. The faceted images are DAMIT sky
  projection models (with artificial lighting) computed for the
  corresponding mean epoch of each image and are shown on the 
  same angular scale (the models have a 12\% uncertainty in 
  scale).}
\label{juno_shifted}
\end{figure*}

The ten images of Juno are shown in Figure~\ref{juno_shifted}.  The
absolute position was measured in each image by computing the centroid
of all pixels above the $5~\sigma$ level, where $\sigma$, the image
rms, was defined by an annulus surrounding the object (see
Table~\ref{tabletwo} notes).  These pixels were weighted uniformly to
avoid influence of surface brightness variations across the face of
the object.  The difference between the centroid position and the
image phase center yields the observed offset from the JPL ephemeris,
which is stable in right ascension at $\approx +60$~mas, but slowly
varying in declination.  The integrated flux density of the source was
measured by integrating over all pixels above the $3~\sigma$ level.
The peak positions, intensities, and the corresponding Planck
brightness temperatures ($T_{\rm B}$) are also listed in
Table~\ref{tabletwo}.  Note that these $T_{\rm B}$ differ from the
Rayleigh-Jeans approximation by $+5.6$~K.  An estimate of the median
brightness temperature across the surface was computed for each image
by finding the median pixel intensity in the clean component model
image and dividing by the solid angle of a pixel.  The result is
typically 10-15~K below the peak $T_{\rm B}$.

% body length in image0 = 131mm, width of raw damit image = 151mm = .24arcsec
% body length in first panel in figure = 45mm, width of panel = 70mm
%  0.24*131/151 * 70/45 = .324 arcsec, which is close to the requested .320
%  so it seems the scale is correct in the figure

% 120 miles

\subsection{Size and shape measurement}
% pole = RA=133deg, dec=47deg

\label{size}

In order to obtain a size measurement independent of the existing
shape models, we consider the underlying source to be an elliptical
disk with uniform brightness.  First, we first fit the observed image
from each epoch with a two-dimensional elliptical Gaussian, recording
the position angle and major and minor axes as the target parameters.
We then create a disk model image using the target parameters, but
with the major and minor axes increased by 30\% to account for the
bulk of the size underestimate resulting from the Gaussian brightness
profile.  We then convolve this disk model image with a 42~mas beam to
match the observations.  Next we iteratively refined the disk model
parameters until an elliptical Gaussian fit to the convolved disk
model image matched the target parameters.  The resulting uniform
brightness elliptical disk model parameters for each epoch are listed
in Table~\ref{tabletwo}.  The geometric mean of the median of the
major and minor axes of the ten elliptical disk models
($0.181\arcsec\pm0.003\arcsec$) corresponds to $d_{\rm mm} =
259\pm4$~km. This mean diameter is consistent with the size derived
from 250~GHz single dish flux density measurements under the
assumption of unity emissivity \citep[$\epsilon_\nu=1$; $d_{\rm
mm}=253.4\pm7.4$;][]{Altenhoff94}.
%A value of $\epsilon_\nu=1$ is also
%consistent with 270~GHz observations of \citet{Redman90}.  
The mean
diameter is also in reasonable agreement with the triaxial geometric
mean ($250\pm5$~km) of the three axes of the \citet{Drummond08}
triaxial ellipsoid model, the equivalent diameter of the
\citet{Durech11} model ($252\pm29$~km), and the effective diameter
measured by radar \citep[$265\pm30$~km;][]{Magri07}.  The {\it IRAS} Minor
Planet Survey (IMPS) inferred a radiometric mean diameter of
$234\pm11$~km, but IMPS diameters are systematically low compared to
occultation diameters \citep{Tedesco04}.  We note that our assumption
of uniform brightness is somewhat unphysical, as the daytime surface
temperature will vary as a function of local hour and latitude in more
realistic thermal models \citep{Lebofsky89}.  We estimate that
our method will yield sizes that are $\approx$2-4\% smaller than the
mean physical size.
Thus, further detailed comparisons of the millimeter images to any 
specific shape model should be performed in the context of a 
thermal model.

\subsection{Comparison to DAMIT shape model}

The model images shown in Figure~\ref{juno_shifted} were obtained
using the DAMIT online tool 
\footnote{\url{http://astro.troja.mff.cuni.cz/projects/asteroids3D/web.php}}
\citep[see also the Interactive Service for Asteroid
Models\footnote{\url{http://isam.astro.amu.edu.pl}}
(ISAM);][]{Marciniak12}.  This tool provides a prediction of the
projected appearance of the asteroid on the sky as viewed from Earth
in equatorial coordinate orientation at any observed Julian date.  The
light travel time effect is taken into account (in this case
16.4~minutes).  The DAMIT prediction for the ESO 1.5~m speckle
observation of \citet{Baier83} is in excellent agreement with the
shape of the observed image, obtained over 34 years ago.  Likewise,
the shape of the DAMIT prediction is in good agreement with most of
the ALMA images.  The images we show here use artificial lighting in
order to show the full geometrical extent of the body, which will emit
millimeter emission that is only mildly modulated by solar
illumination.  In a qualitative sense, it is perhaps images 4 and 5
that are most discrepant from the model in terms of ellipticity and
orientation. As for angular scale, the quoted accuracy of the DAMIT
images is $\pm12$\%.  The ALMA-derived major axes
(Table~\ref{tabletwo}) are systematically $\approx6$\% smaller than
the maximum extent in the DAMIT images, however much of this
difference could be due to the simplistic model used in \S~\ref{size}.

\subsection{Surface features}

\begin{figure*}
\plotone{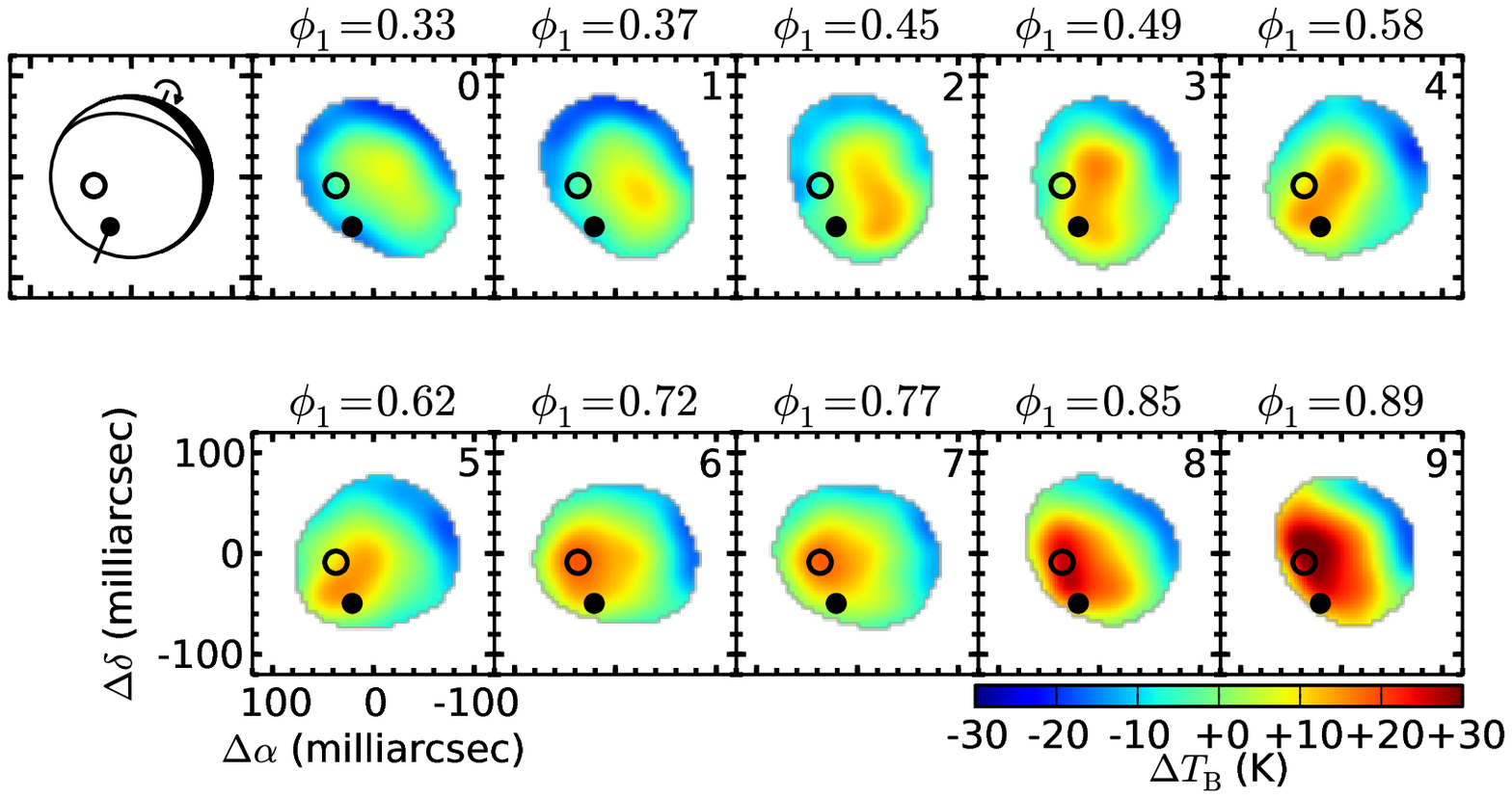}    % tenpaneldiff_individual_shifts.eps
\caption[]{Residual images of Planck brightness temperature ($T_{\rm B}$)
in equatorial offset coordinates created by removing a model image 
of uniform brightness from the
images in Figure~\ref{juno_shifted}.  The open circle indicates the location
of the subsolar point.  The color scale range is $\pm30$~K with
respect to the median values of $T_{\rm B}$ in column~8 of
Table~\ref{tabletwo}.  The south pole drawn is the same as in
Figure~\ref{orbitdiagram}, and the sense of rotation is clockwise.
The different position angle here is due to the combination of the
different coordinate system, the foreshortening effect of the heavily
inclined pole, and the significant heliocentric latitude.
\\
\\
\\
  }
\label{juno_diff}
\end{figure*}

To accentuate the variation of brightness across the object, we have
created a model image of uniform brightness corresponding to each of
the ten images.  We begin with the clean component model image, and
compute the median value within the central 0.15\arcsec\/ diameter
(\S~\ref{images}).  We then place this value into all pixels of the
clean model that are inside the 40\% level in the clean image, and
place zero elsewhere.  This approach leads to a model image comparable
to the angular dimensions of Juno.  We then convolve this model with
the 42~mas beam and subtract it from the clean image.  This
subtraction enables the identification of areas of lower vs. higher
brightness temperature, regardless of the relative calibration
accuracy between the different images.  The results are shown in
Figure~\ref{juno_diff}.  In most of the frames there is a consistent
pattern of the northwest edge being the coolest portion, which matches
the location of the evening terminator.  Also, the warmest part of the
image appears correlated with the subsolar point.  In the first five
images, the afternoon area following the subsolar point is the warmest
point. Meanwhile, in the last four images, the warmest point is at (or
very close to) the subsolar point. The temperature contrast becomes
particularly pronounced in the final two images as the surface median
$T_{\rm B}$ declines (Table~\ref{tabletwo}).

\section{Discussion}
\label{discuss}
In the simple equilibrium model, the expected disk-averaged brightness
temperature, $T_{B}(\nu)$, of an asteroid is determined by its
spectral emissivity, $\epsilon_\nu$, and its mean equilibrium
temperature, $T_{eq}$:
\begin{equation}
 T_{B}(\nu) = \epsilon_\nu T_{eq}
\end{equation}
\begin{equation} 
   T_{eq} = f(1-A)^{1/4}r^{-1/2},
\end{equation}
where $A$ is its bolometric Bond albedo, $r$ is its heliocentric
distance in AU, and $f=329$~K for non-rotating objects and 277~K for
fast-rotating objects \citep{Cremonese02,Kellermann66}.  The
bolometric Bond albedo is a product of the bolometric geometric albedo
($p$) and the bolometric phase integral ($q$). In principle, both of
these components represent integrals of wavelength-dependent
quantities which must be measured and weighted by the solar flux
spectrum \citep{Hansen77}, but they are often approximated as being
wavelength independent.  Measurements of $p$ for Juno at optical
wavelengths ($p_{\rm V}$) range from $\approx0.13$
\citep{Hansen77,Brown82} to 0.15 \citep{Morrison77,Zellner77}, while a
value of $\approx0.22$ has been found at mid-infrared wavelengths
\citep{Ryan10,Tedesco04}.
% By comparison, the radar reflectivity ($\hat\sigma_{\rm OC}$) is
% $0.14\pm0.05$ \citep{Magri07}.
% , which is equal to the mean such value of S-class asteroids .  
The observed variation in optical brightness versus phase angle yields
a slope parameter $G = 0.17\pm0.03$ \citep{Lagerkvist92}, which in turn
yields $q_{\rm V} = 0.41\pm0.02$ from the relation of \citet{Bowell89}.  
% Multiplying $q_{\rm V}$
% by the range of values for $p_{\rm V}$ yields $A = 0.05-0.09$.
Combining $q_{\rm V}$ with the more recent measurements of $p_{\rm
IR}$, we will proceed with $A=0.09$, which matches the result
others have obtained by using the $p_{\rm V}$ values along
with $q_{\rm V}=0.6$ appropriate for the Moon.  In any case, the
dependence of $T_{eq}$ on $A$ is quite mild.  Assuming
$\epsilon_\nu=1$, the equilibrium model prediction for $T_{B}$ for our
Juno observations ($r=2.072$~AU) is then 188~K to 223~K, depending on
$f$.  Juno is a fast rotator, but its polar axis was pointed
significantly in line with the Sun during the ALMA observations (Fig.~1).
Therefore, an $f$ value in the lower half of the range is likely to be
appropriate, which is consistent with our measured median $T_{\rm
B}^{\rm obs}$(233~GHz) of $197\pm15$~K.  A previous single-dish
estimate of $T_{\rm B}^{\rm obs}$(345~GHz) was $146\pm48$~K at
$r=3.132$~AU \citep{Chamberlain09}, which scales to $180\pm59$~K at
our smaller value of $r$.

Moving beyond the equilibrium model, we next consider the Standard
Thermal Model for asteroids \citep[STM;][]{Lebofsky89}.  Because ALMA
resolves the surface of Juno, we can compare the observed peak
brightness temperature with the expected subsolar surface temperature
($T_{\rm ss}$) in the STM.  For Juno's values of $A$ and $r$, and its
beaming factor of $\eta = 0.76$ \citep{Spencer89}, the expected
$T_{\rm ss}$ is 286~K.  This value is significantly higher than the
peak $T_{\rm B}$ of $222\pm13$~K observed by ALMA, in contrast to
previous mid-infrared measurements of $T_{\rm ss}$ on Juno which are
consistent with the STM prediction \citep{Lim05}.  If interpreted in
the context of a model where all of the emission arises from the
surface, such a discrepancy could be interpreted as an
``effective'' $\epsilon_\nu \sim 0.8$ at 1.3~mm.
Effective emissivity is a quantity which can encompass many effects
besides the physical emissivity of the material, including sub-surface
sounding of deeper colder layers \citep{Fornasier13,Redman98}.
%, as it has been, for example, for (disk-averaged) {\it
%Herschel} observations of Transneptunian objects 
Indeed, in terms of a physical model with temperature gradients in the
sub-surface material of up to 50~K~mm$^{-1}$, such a large temperature
discrepancy is expected to be seen in millimeter wavelength emission,
which arises from material at a range of depths, even when the bulk
material has $\epsilon_\nu \approx 1$ \citep[see, e.g.,][]{Keihm13}.

To interpret the enhanced brightness temperature of the subsolar point
in most of the Juno images, we consider heat conduction in the
regolith following the equations in \citet{Lagerros96}.  The thermal
inertia ($\Gamma$) is a function of the surface material density
($\rho$), thermal conductivity ($\kappa$), and specific heat capacity
($c_{\rm p}$) of the soil, while the thermal skin depth ($l_{\rm s}$)
also depends on the angular rotation rate ($\omega$),  
\begin{equation}
\Gamma    = \sqrt{\kappa \rho c_{\rm p}}
\end{equation}
\begin{equation}
l_{\rm s} = \sqrt{\frac{\kappa}{\rho c_{\rm p} \omega}} = \frac{\kappa}{\Gamma \sqrt{\omega}}. 
\end{equation}
Using the surface density of Vesta derived from its radar albedo
\citep[1.75~g~cm$^{-3}$;][]{Chamberlain09}, 
along with values for $\kappa$ and $c_{\rm p}$ in the porous lunar 
surface layer
\citep[$2\times10^{-5}$ W~cm$^{-1}$~K$^{-1}$ and
0.6~J~g$^{-1}$~K$^{-1}$, respectively;][]{Keihm84}, yields $\Gamma =
46$~J~m$^{-2}$~s$^{-0.5}$~K$^{-1}$
% \citep[similar to the value of 50 assumed by][]{Lagerros98} 
and $l_{\rm s} = 2.8$~mm.
% and the corresponding thermal
% parameter, $\Theta$, is 0.65 (for $\epsilon_\nu=1$).  
Since $l_{\rm s}$ is only 2.2 wavelengths, we can expect the observed
continuum emission to arise from a mix of solar-heated surface
material and deeper unheated material.
Thus, the correlation of Juno's brightness temperature with the
subsolar point is not surprising.  However, the fact that the
brightest point moves from the subsolar afternoon area in the first
half of the images to near coincidence with the subsolar point in the
latter half of the images, suggests a change in the soil properties
with longitude.  For example, in the case that $\kappa$ is constant
with depth and longitude, $l_{\rm s}$ would scale inversely with
$\Gamma$ due to changes in either $\rho$ or $c_{\rm p}$.  Thus, if
soil with relatively lower values of $\rho$ or $c_{\rm p}$ exists at
these longitudes ($\phi_1 \sim 0.7-0.9$), it could have a lower
thermal inertia, deeper skin depth, and consequently a greater
proportion of the millimeter emission arising from heated material.  A
lower inertia across most of this side of Juno might also explain the
lower median $T_{\rm B}$ observed, as cooling would proceed more
rapidly as the angle from the subsolar point increases.  In any case,
variations in thermal inertia across the surface of an asteroid are
quite plausible, particularly in light of the detailed variations on
Vesta reported from the Visual and Infrared mapping spectrometer on
the {\it Dawn} spacecraft \citep{Capria14}.

Recently, two asteroids have been measured at 0.53 and 1.6~mm by the
Microwave Instrument for the Rosetta Orbiter (MIRO) during close
encounters by the European Space Agency {\it Rosetta} spacecraft: the
small ($\sim$6~km) object {\v S}teins \citep{Gulkis10}, and the larger
($\sim$100~km) object Lutetia \citep{Gulkis12}.  In contrast to {\v
S}teins, which has a high, rock-like inertia
($\Gamma=450-850$~J~m$^{-2}$~s$^{-0.5}$~K$^{-1}$) and
$\epsilon_\nu=0.85-0.9$, Lutetia exhibits a very low
thermal inertia ($\Gamma=20$~J~m$^{-2}$~s$^{-0.5}$~K$^{-1}$) in the
upper 1-3~cm much like the fine dust of the lunar regolith, with an
emissivity consistent with reflection from a surface with dielectric
constant of 2.3 ($\epsilon_\nu = 0.958$).  These latter properties may
be similar to what ALMA has seen on Juno, particularly in the latter
half of the images.  This result is perhaps not surprising in
the context of the subsequent thermophysical modeling of
\citet{Keihm13}, which finds that low thermal inertias and
$\epsilon_\nu \sim 1$ can fit the infrared to centimeter spectral
energy distributions of the asteroids Ceres, Vesta, Pallas, and
Hygiea.

The possibility of a recent impact on Juno was raised by the detection
of a region of reduced 934~nm brightness in AO images correlated with
a spatial ``bite'' feature on the limb \citep{Baliunas03}.
Unfortunately, the north/south orientation of the \citet{Baliunas03}
images is not specified.  However, judging from the DAMIT model images
at that epoch, the AO images would appear to be oriented with south up
because in this case, a depression in the model images would map
closely to the proposed crater near the limb of the fifth AO image.
If so, then the crater is located near the north pole. As shown in
Figure~\ref{orbitdiagram}, the ALMA viewing angle of Juno differs by
114\arcdeg\/ from the \citet{Baliunas03} AO images such that the north
pole is not visible.  On the other hand, if the AO image is oriented
with north up, as it is in similarly-acquired images of Vesta by a
subset of these authors \citep{Shelton97}, then the crater would lie
between the equator and the south pole, placing it at a latitude that
crosses near the center of the ALMA view.  The rotational phase of the
fifth AO image ($\phi_2 = 0.29$) is close to that of ALMA image 3,
thus a feature on the limb in the AO image would cross the sub-Earth
point in the ALMA image 114\arcdeg+90\arcdeg=204\arcdeg\/ later
(i.e. at $\phi_2 = 0.86$).  This phase corresponds to ALMA images 8
and 9.  Those images do show the highest temperature contrast with
respect to the subsolar point, which could be consistent with a lower
thermal inertia in the excavated material surrounding the crater.
Clearly, future ALMA observations of complete rotations of Juno,
preferably at multiple phases and wavelengths, will be necessary to
explore this phenomenon further and develop an accurate, full-surface
thermophysical model.

\section{Conclusions}

Our ALMA long-baseline observations of Juno provide the first
ground-based images that significantly resolve the surface of an
asteroid at millimeter wavelengths.  They provide an independent set
of size and shape measurements which confirm our current knowledge
expressed by the DAMIT and triaxial ellipsoid models.  Future ALMA
observations of main belt asteroids, including both
spatially-unresolved photometric lightcurves
\citep[e.g.,][]{Moullet10} and resolved images, can be used to test
and refine the existing three-dimensional models.  We note that ALMA
can potentially achieve significantly higher physical resolution on
Juno that these initial observations offer.  For example, a factor of
three improvement (to 20~km resolution) would be possible by observing
in the 345~GHz band with a similar antenna configuration at a future
favorable opposition (e.g., 16 Nov 2018, $\Delta=1.04$~AU).  These
observations would match the resolution of MIRO's 1.6~mm channel
during {\it Rosetta's} flyby of Lutetia.  At these scales, measurements 
of the brightness temperature will provide new information about the
surfaces of these bodies.
%, as well as planetary
% moons and Kuiper Belt Objects \citep[][]{Moullet11}.  
To develop accurate thermophysical models, it will be important to
observe them at multiple (sub)millimeter wavelengths where a drop in
emissivity to values of $\sim0.6-0.8$ has been reported
\citep[e.g.,][]{Mueller98} particularly for rockier bodies
\citep[e.g.,][]{Gulkis10}.  ALMA can also potentially measure the
mutual orbit of smaller binary asteroids, providing important
information on the mass and density of such objects
\citep[e.g.,][]{Carry15}.  Finally, the ability of ALMA to deliver
very accurate astrometry will enable better long-term modeling of
asteroid orbits, leading to improved predictions \citep{Busch09}.

\acknowledgments

This paper makes use of the following ALMA data set:
\dataset{ADS/JAO.ALMA\#2011.0.00013.SV}.  ALMA is a partnership of ESO
(representing its member states), NSF (USA) and NINS (Japan), together
with NRC (Canada), NSC and ASIAA (Taiwan), and KASI (Republic of
Korea), in cooperation with the Republic of Chile.  The Joint ALMA
Observatory is operated by ESO, AUI/NRAO and NAOJ.  The National Radio
Astronomy Observatory is a facility of the National Science Foundation
operated under cooperative agreement by Associated Universities, Inc.
This research has made use of NASA's Astrophysics Data System.  We
thank Thomas M{\"u}ller, Mark Gurwell, Bryan Butler, Rafael Hiriart,
Ralph Marson, Dirk Petry, and Vivek Dhawan for useful discussions.

{\it Facilities:}\facility{ALMA}.

%   1.495978707e8 * 0.266 / 206264.8 = 192.9 km
% Barnard 1900
%             Obs    JPL    AU   @1AU
% 1900-09-13  0.26   0.2690 1.199 .224
% 1900-09-19  0.23   0.276  1.17  .236 
% 1900-09-29  0.23   0.282  1.14  .247
% 1900-10-08  0.22   0.283  1.14  .248
% 1900-10-11  0.20   0.282  1.14  .247


\begin{thebibliography}{}

\bibitem[ALMA partnership et al.(2015)]{ALMA15a}
ALMA partnership, Fomalont, E., Vlahakis, C., et al.\ 2015a, \apjl, submitted

\bibitem[Altenhoff et al.(1994)]{Altenhoff94} 
Altenhoff, W.~J., Johnston, K.~J., Stumpff, P., \& Webster, W.~J.\ 1994, 
\aap, 287, 641 

\bibitem[Baier \& Weigelt(1983)]{Baier83} Baier, G., \& Weigelt,
  G.\ 1983, \aap, 121, 137

\bibitem[Baliunas et al.(2003)]{Baliunas03} Baliunas, S., Donahue, 
R., Rampino, M.~R., et al.\ 2003, \icarus, 163, 135   % 0.022''/pixel

\bibitem[Barnard(1900)]{Barnard00} Barnard, E.~E.\ 1900, \mnras, 
61, 68 

\bibitem[Barnard(1895)]{Barnard95} Barnard, E.~E.\ 1895, \mnras, 
56, 55 

\bibitem[Birch \& Taylor(1989)]{Birch89}   % old measurement of rotation period
Birch, P.~V., \& Taylor, R.~C.\ 1989, \aaps, 81, 409 

\bibitem[Bowell et al.(1989)]{Bowell89} Bowell, E., Hapke, B., 
Domingue, D., et al.\ 1989, Asteroids II, , ed. R. Binzel,
University of Arizona Press, 524 

\bibitem[Brown et al.(1982)]{Brown82} Brown, R.~H., Morrison, 
D., Telesco, C.~M., \& Brunk, W.~E.\ 1982, \icarus, 52, 188 

\bibitem[Bus \& Binzel(2002)]{Bus02} 
Bus, S.~J., \& Binzel, R.~P.\ 2002, \icarus, 158, 106 

\bibitem[Busch(2009)]{Busch09} Busch, M.~W.\ 2009, \icarus, 
200, 347 

% 35 GHz measurements of absorption in rock powder
\bibitem[Campbell \& Ulrichs(1969)]{Campbell69}   
Campbell, M.~J., \& Ulrichs, J.\ 1969, \jgr, 74, 5867 

\bibitem[Capria et al.(2014)]{Capria14} Capria, M.~T., Tosi, F., 
De Sanctis, M.~C., et al.\ 2014, \grl, 41, 1438 

\bibitem[Carry et al.(2015)]{Carry15} Carry, B., Matter, A., 
Scheirich, P., et al.\ 2015, \icarus, 248, 516 

% 2.7E19 kg = 1.37E-11 Msun, density = 3.68
%\bibitem[Carry(2012)]{Carry12} 
% Carry, B.\ 2012, \planss, 73, 98 

% speckle imaging, but perhaps an instrumental problem because no evidence
% for an elliptical shape is seen, but DAMIT model is clearly elliptical
% on JD=2451817.447916666
%\bibitem[Cellino et al.(2003)]{Cellino03} Cellino, A., Diolaiti,   
%E., Ragazzoni, R., et al.\ 2003, \icarus, 162, 278 

\bibitem[Chamberlain et al.(2009)]{Chamberlain09} Chamberlain, M.~A., 
Lovell, A.~J., \& Sykes, M.~V.\ 2009, \icarus, 202, 487 

\bibitem[Chamberlain et al.(2007)]{Chamberlain07} Chamberlain, M.~A., 
Lovell, A.~J., \& Sykes, M.~V.\ 2007, \icarus, 192, 448 

\bibitem[Chapman et al.(1975)]{Chapman75} Chapman, C.~R., 
Morrison, D., \& Zellner, B.\ 1975, \icarus, 25, 104 

\bibitem[Cowling(1984)]{Cowling84} Cowling, S.~A.\ 1984, \mnras, 
209, 415 

\bibitem[Cremonese et al.(2002)]{Cremonese02} Cremonese, G., 
Marzari, F., Burigana, C., \& Maris, M.\ 2002, \na, 7, 483 

%\bibitem[Cunningham(2004)]{Cunningham04} Cunningham, C.~J.\ 2004, 
%Journal of Astronomical History and Heritage, 7, 116 

\bibitem[Degewij et al.(1979)]{Degewij79} Degewij, J., Tedesco, 
E.~F., \& Zellner, B.\ 1979, \icarus, 40, 364 

\bibitem[Dollfus(1971)]{Dollfus71} Dollfus, A.\ 1971, Proc. of IAU
Colloquium 12, NASA Special Publication, 267, 25

\bibitem[Dotto et al.(2002)]{Dotto02} Dotto, E., Barucci, M.~A.,
M{\"u}ller, T.~G., Storrs, A.~D., \& Tanga, P.\ 2002, Asteroids III,
University of Arizona Press, 219

\bibitem[Dotto et al.(2000)]{Dotto00} 
Dotto, E., M{\"u}ller, T.~G., Barucci, M.~A., et al.\ 2000, \aap, 358, 1133 

\bibitem[Dotto et al.(1995)]{Dotto95} Dotto, E., De Angelis, 
G., Di Martino, M., et al.\ 1995, \icarus, 117, 313 

\bibitem[Drummond \& Christou(2008)]{Drummond08} % Lick IR AO 0.076''/pixel
Drummond, J., \& Christou, J.\ 2008, \icarus, 197, 480 

\bibitem[{\v D}urech et al.(2011)]{Durech11} {\v D}urech, J., 
Kaasalainen, M., Herald, D., et al.\ 2011, \icarus, 214, 652 

\bibitem[{\v D}urech et al.(2010)]{Durech10} 
{\v D}urech, J., Sidorin, V., \& Kaasalainen, M.\ 2010, \aap, 513, AA46 

%\bibitem[Feierberg et al.(1982)]{Feierberg82} Feierberg, M.~A., 
%Larson, H.~P., \& Chapman, C.~R.\ 1982, \apj, 257, 361 

%\bibitem[Fomalont et al.(2014)]{Fomalont14} Fomalont, E., van 
%Kempen, T., Kneissl, R., et al.\ 2014, The Messenger, 155, 19

\bibitem[Fornasier et al.(2013)]{Fornasier13} 
Fornasier, S., Lellouch, E., M{\"u}ller, T., et al.\ 2013, \aap, 555, AA15 

\bibitem[Gaffey et al.(1993a)]{Gaffey93a} Gaffey, M.~J., Burbine, 
T.~H., \& Binzel, R.~P.\ 1993a, Meteoritics, 28, 161 

%\bibitem[Gaffey et al.(1993b)]{Gaffey93b} Gaffey, M.~J., Burbine, 
%T.~H., Piatek, J.~L., et al.\ 1993b, \icarus, 106, 573 

%\bibitem[Giorgini et al.(1996)]{Giorgini96} 
%Giorgini, J.~D., Yeomans, D.~K., Chamberlin, A.~B., et al.\ 1996, Bulletin of the American 
%Astronomical Society, 28, 1158 

% \bibitem[Giorgini et al.(2001)]{2001DPS....33.5813G} Giorgini, J.~D., 
% Chodas, P.~W., %\& Yeomans, D.~K.\ 2001, Bulletin of the American 
% Astronomical Society, 33, 1562 

\bibitem[Gulkis et al.(2012)]{Gulkis12} 
Gulkis, S., Keihm, S., Kamp, L., et al.\ 2012, \planss, 66, 31 

\bibitem[Gulkis et al.(2010)]{Gulkis10} 
Gulkis, S., Keihm, S., Kamp, L., et al.\ 2010, \planss, 58, 1077 

% \bibitem[Gulkis et al.(2007)]{Gulkis07} Gulkis, S., Frerking,   % MIRO description
% M., Crovisier, J., et al.\ 2007, \ssr, 128, 561 

\bibitem[Hansen(1977)]{Hansen77} Hansen, O.~L.\ 1977, \icarus, 
31, 456 

\bibitem[Hezaveh et al.(2013)]{Hezaveh13} Hezaveh, Y.~D., 
Marrone, D.~P., Fassnacht, C.~D., et al.\ 2013, \apj, 767, 132 

\bibitem[Hills et al.(2010)]{Hills10} Hills, R.~E., Kurz, 
R.~J., \& Peck, A.~B.\ 2010, \procspie, 7733, 773317 

\bibitem[Kaasalainen et al.(2002)]{Kaas02} Kaasalainen, M.,
  Torppa, J., \& Piironen, J.\ 2002, \icarus, 159, 369

\bibitem[Keihm et al.(2013)]{Keihm13} Keihm, S., Kamp, L., 
Gulkis, S., et al.\ 2013, \icarus, 226, 1086 

\bibitem[Keihm(1984)]{Keihm84} 
Keihm, S.~J.\ 1984, \icarus, 60, 568 

\bibitem[Kellermann(1966)]{Kellermann66} Kellermann, K.~I.\ 1966, 
\icarus, 5, 478 

\bibitem[Lagerkvist et al.(1992)]{Lagerkvist92} Lagerkvist, C.-I.,
Magnusson, P., Williams, I.~P., et al.\ 1992, \aaps, 94, 43

%\bibitem[Lagerros(1998)]{Lagerros98} Lagerros, J.~S.~V.\ 1998, \aap,
%332, 1123

\bibitem[Lagerros(1996)]{Lagerros96}   % skin depth equationn
Lagerros, J.~S.~V.\ 1996, \aap, 310, 1011 

\bibitem[Lanyi et al.(2010)]{Lanyi10} Lanyi, G.~E., Boboltz, 
D.~A., Charlot, P., et al.\ 2010, \aj, 139, 1695 

\bibitem[Lebofsky \& Spencer(1989)]{Lebofsky89} 
Lebofsky, L.~A., \& Spencer, J.~R.\ 1989, in Asteroids II, ed. R. Binzel,
University of Arizona Press, 128 

%\bibitem[Lebofsky et al.(1986)]{Lebofsky86} Lebofsky, L.~A., 
%Sykes, M.~V., Tedesco, E.~F., et al.\ 1986, \icarus, 68, 239 

\bibitem[Lim et al.(2005)]{Lim05} Lim, L.~F., McConnochie, 
T.~H., Bell, J.~F., \& Hayward, T.~L.\ 2005, \icarus, 173, 385 

\bibitem[Lovell(2008)]{Lovell08} Lovell, A.~J.\ 2008, \apss, 313, 191 

\bibitem[Magnusson(1986)]{Magnusson86} Magnusson, P.\ 1986, \icarus,
  68, 1

\bibitem[Magri et al.(2007)]{Magri07} Magri, C., Nolan, M.~C., 
Ostro, S.~J., \& Giorgini, J.~D.\ 2007, \icarus, 186, 126 

\bibitem[Marciniak et al.(2012)]{Marciniak12}  
Marciniak, A., Bartczak, P., Santana-Ros, T., et al.\ 2012, \aap, 545, AA131 

% \bibitem[Masiero et al.(2014)]{Masiero14} Masiero, J.~R., Grav, 
% T., Mainzer, A.~K., et al.\ 2014, \apj, 791, 121 

\bibitem[Millis et al.(1981)]{Millis81} Millis, R.~L., 
Wasserman, L.~H., Bowell, E., et al.\ 1981, \aj, 86, 306 

%\bibitem[Mitchell et al.(1996)]{Mitchell96}   % Ceres/Vesta/Pallas radar, giving densities
%Mitchell, D.~L.,  Ostro, S.~J., Hudson, R.~S., et al.\ 1996, \icarus, 124, 113 

\bibitem[Morrison(1977)]{Morrison77}   % 0.151
Morrison, D.\ 1977, \icarus, 31, 185 

%\bibitem[Morrison(1974)]{Morrison74} Morrison, D.\ 1974, \apj,  % 0.15, D=252km
%194, 203 

%\bibitem[Moullet et al.(2011)]{Moullet11} Moullet, A., Lellouch, 
%E., Moreno, R., \& Gurwell, M.\ 2011, \icarus, 213, 382 

\bibitem[Moullet et al.(2010)]{Moullet10} 
Moullet, A., Gurwell, M., \& Carry, B.\ 2010, \aap, 516, L10 

%\bibitem[M{\"u}ller \& Barnes(2007)]{Mueller07} 
%M{\"u}ller, T.~G., \& Barnes, P.~J.\ 2007, \aap, 467, 737 

\bibitem[M{\"u}ller \& Lagerros(1998)]{Mueller98} 
M{\"u}ller, T.~G., \& Lagerros, J.~S.~V.\ 1998, \aap, 338, 340

\bibitem[Rau \& Cornwell(2011)]{Rau11} Rau, U., \& Cornwell,
  T.~J.\ 2011, \aap, 532, AA71

%\bibitem[Redman et al.(1990)]{Redman90} Redman, R.~O., Feldman, 
%P.~A., Halliday, I., 
%\& Matthews, H.~E.\ 1990, Asteroids, Comets, Meteors III, 163 

% \bibitem[Redman et al.(1992)]{Redman92} Redman, R.~O., Feldman, 
% P.~A., Matthews, H.~E., Halliday, I., \& Creutzberg, F.\ 1992, \aj, 104, 405 

\bibitem[Redman et al.(1998)]{Redman98} Redman, R.~O., Feldman, 
P.~A., \& Matthews, H.~E.\ 1998, \aj, 116, 1478 

%\bibitem[Russell(1916)]{Russell16} Russell, H.~N.\ 1916, \apj, 
%43, 173 

\bibitem[Ryan \& Woodward(2010)]{Ryan10} 
Ryan, E.~L., \& Woodward, C.~E.\ 2010, \aj, 140, 933 

\bibitem[Schroll et al.(1981)]{Schroll81} 
Schroll, A., Schober, H.~J., \& Lagerkvist, C.~I.\ 1981, \aap, 104, 296 

\bibitem[Shelton et al.(1997)]{Shelton97} Shelton, J.~C., 
Schneider, T., \& Baliunas, S.~L.\ 1997, \procspie, 3126, 321 

\bibitem[Shinokawa et al.(2002)]{Shinokawa02} Shinokawa, K., 
Takahashi, S., Ogawa, K., et al.\ 2002, \memsai, 73, 658 

\bibitem[Spencer et al.(1989)]{Spencer89} 
Spencer, J.~R., Lebofsky, L.~A., \& Sykes, M.~V.\ 1989, \icarus, 78, 337 

\bibitem[Takahashi et al.(2009)]{Takahashi09} 
Takahashi, S., Yoshida, F., Shinokawa, K., Mukai, T., 
\& Kawabata, K.~S.\ 2009, \aj, 138, 951 

\bibitem[Tedesco et al.(2004)]{Tedesco04} Tedesco, E.~F., Noah, P.~V.,
  Noah, M., \& Price, S.~D.\  {\it IRAS} Minor Planet
  Survey, NASA Planetary Data System, 2004

%\bibitem[Tholen(1989)]{Tholen89} Tholen, D.~J.\ 1989, in Asteroids 
%II, ed. R. Binzel, University of Arizona Press, 1139 

\bibitem[van Kempen et al.(2014)]{vanKempen14} van Kempen, T., Kneissl, R.,
Marcelino, N., et al., ALMA Memo 599

\bibitem[Viikinkoski et al.(2015)]{Viikinkoski15} Viikinkoski, M., 
Kaasalainen, M., \& Durech, J.\ 2015, arXiv:1501.05958 

% \bibitem[Webster \& Johnston(1989)]{Webster89} 
% Webster, W.~J., Jr., \& Johnston, K.~J.\ 1989, \pasp, 101, 122 

\bibitem[Zellner et al.(1977)]{Zellner77} 
Zellner, B., Leake, M., Lebertre, T., Duseaux, M., \& Dollfus, A.\ 1977, 
Lunar and Planetary Science Conference Proceedings, 8, 1091 

%\bibitem[Baer \& Chesley(2008)]{Baer08} % 1.4-2.0E-11 Msun, references therein
% Baer, J., \& Chesley, S.~R.\ 2008, Celestial Mechanics and Dynamical 
% Astronomy, 100, 27 
%

\end{thebibliography}
\end{document}